# Signature of band inversion in the perovskite thin-film alloys BaSn$_{1-x}$Pb$_x$O$_3$

Junichi Shiogai[*], Takumaru Chida, Kenichiro Hashimoto[†],

Kohei Fujiwara, Takahiko Sasaki, and Atsushi Tsukazaki

*Institute for Materials Research, Tohoku University, Sendai 980-8577, Japan*

[*]E-mail: junichi.shiogai@imr.tohoku.ac.jp

[†] *Present address: Department of Advanced Materials Science, The University of Tokyo,*

*Kashiwa, 277-8561, Japan*

**Abstract**

Perovskite oxides $AB$O$_3$ containing heavy $B$-site elements are a class of candidate materials to host topological metals with a large spin-orbit interaction. In contrast to the band insulator BaSnO$_3$, the semimetal BaPbO$_3$ is proposed to be a typical example with an inverted band structure, the conduction band of which is composed of mainly the O-2$p$ orbital. In this study, we exemplify a band-gap modification by systematic structural, optical, and transport measurements in BaSn$_{1-x}$Pb$_x$O$_3$ films. A sudden suppression of the conductivity and an enhancement of the weak antilocalization effect at $x$ = 0.9 indicate the presence of a singular point in the electronic structure as a signature of the band inversion. Our findings provide an intriguing platform for combining topological aspects and electron correlation in perovskite oxides based on band-gap engineering.

**Main text**

The spin-orbit interaction (SOI) plays a crucially important role in topological quantum materials for modification of the electronic band structure accompanied by band inversion [1,2]. The band inversion occurs when the band gap closes at a critical point through continuous deformation of the electronic band structure as functions of various control parameters, such as chemical composition [3,4,5,6,7], pressure [8], strain [9], and heterostructure architecture [10,11]. For instance, topological phase transitions (TPTs) have been reported to occur by chemical doping in topological insulators such as Bi-Sb binary alloys [3,4], TlBi(Te,Se)$_2$ [5], and (Bi,In)$_2$Se$_3$ [6,7], and in the topological crystalline insulator (Pb,Sn)Te [8]. Along with TPT, topological surface states or three-dimensional linearly dispersed bands appear in these topological materials [12], hosting exotic topological phenomena in the bulk and heterojunctions [13]. Particularly in pyrochlore Ir oxides, in the presence of electron correlation and a large SOI, exotic quantum phases such as Weyl or Dirac semimetals are predicted to emerge [14,15]. To date, however, only a very limited number of experiments has been reported to verify such topological features in oxide materials [16,17] and thin-film heterostructures [18].

Perovskite oxides $AB$O$_3$, the $B$ site of which is constituted by 3$d$ transition-metal elements, host a wide range of physical properties of strongly correlated electrons such as superconductivity, magnetism [19] and multiferroicity [20]. In these materials, atomic orbitals of $B$-site elements often construct electronic bands around the Fermi energy ($E_F$). From the viewpoint of SOI, BaBiO$_3$ and BaPbO$_3$ are theoretically proposed to host topological phenomena with the band inversion [21,22], which so far have been intensively studied as parent compounds of copper-free high-$T_c$ superconductors despite their relatively low carrier concentration [23,24,25,26]. In this study, we focus on the perovskite oxide thin films with Pb or Sn as *B* site, possibly providing a large atomic SOI. First-principles calculations on BaPbO$_3$ predict that a SOI induces a band inversion between O-2$p$ and Pb-6$s$ orbitals, corresponding to a topological metal with Dirac surface states [22,26]. In the case of narrow gap semiconductors such as InAs, the strength of SOI is characterized to be inversely proportional to the band-gap size by a simple formula [27]. Applying this trend to the perovskite oxides $AB$O$_3$, the band-gap engineering toward materialization of a narrow gap electronic structure by chemical doping is expected to induce a further enhancement of the SOI.

In this study, the theoretically proposed band inversion in BaPbO$_3$, the conduction band of which is mainly composed of the O-2$p$ orbital, has been exemplified by systematic composition-dependent optical and electrical transport measurements in BaSn$_{1-x}$Pb$_x$O$_3$ thin films. Since the valences of Sn$^{4+}$ in BaSnO$_3$ and Pb$^{4+}$ in BaPbO$_3$ are equivalent, the BaSn$_{1-x}$Pb$_x$O$_3$ solid-solution alloy is expected to be a platform for investigation of a TPT in the perovskite structure, holding the *B*-site valence. Here, we depict a conceptual schematic of the electronic band structure modified by alloying BaSnO$_3$ and BaPbO$_3$ (see Figs. 1(a) to 1(d)). One end member BaSnO$_3$ is known to be a trivial insulator with a wide band gap of 3.1 eV as shown in Fig. 1(a) and possesses a cubic perovskite structure (space group: *cP5*) [28,29]. Based on the first-principles calculations, the conduction band is mainly composed of the Sn-5$s$ orbital and the valence band is dominated by the O-2$p$ orbital at the Γ point [29]. In contrast, the other end material BaPbO$_3$ forms an orthorhombic structure with semimetallic nature; band structure calculations reveal that the Pb-6$s$ band is pushed downward below the O-2$p$ band by 0.5 eV at the Γ point, as depicted in Fig. 1(d), because of a strong SOI of the Pb element [22,26]. Therefore, in the BaSn$_{1-x}$Pb$_x$O$_3$ alloy, at a certain Pb composition $x$, electronic band modification is expected to occur accompanied by a band touching. The band touching leads to a low density of states (DOS), which can be observed as an enhanced resistivity due to a reduction of the conducting charge-carrier density in electrical transport measurements and as a systematic variation in optical spectra for $BaSn_{1-x}Pb_xO_3$ films. The experimental observation of these trends can be a definitive signature for the band inversion of $BaPbO_3$.

The 18-20-nm-thick $BaSn_{1-x}Pb_xO_3$ thin films were grown on $SrTiO_3(001)$ substrates by pulsed-laser deposition (PLD) under an oxygen pressure of 0.2 Torr. The substrate temperature was optimized for each Pb composition $x$ between 575 and 350 ºC to suppress reevaporation of Pb. By performing the chemical composition analysis using energy dispersive x-ray spectrometry and inductively coupled plasma atomic emission spectrometry, we have confirmed that the film composition is consistent with the PLD target composition $x$ (see Fig. S1 in Supplemental Material (SM) [30]). The x-ray-diffraction measurements revealed that a $c$-axis oriented perovskite structure was synthesized at all $0 < x < 1$ without secondary phases as shown in Figs. S2 and S3 in SM [30]. The $x$ dependences of the lattice parameters are plotted in Fig. 1(e) for the $BaSn_{1-x}Pb_xO_3$ thin films. The $BaSnO_3$ thin film ($x = 0$) has a cubic perovskite structure with $4.135 \pm 0.004$ Å, nearly consistent with the bulk value ($a = 4.1155$ Å [31]). The cubic structure maintains in $x$ less than 0.42, where the in-plane $a$- and $b$-axis (green squares) and out-of-plane $c$-axis (red circles) lattice constants match well. With further increase of $x$, the $a$ and $b$ axes elongate from the $c$-axis length, indicating that a structural phase transition from cubic to orthorhombic structures occurs. The lattice parameters for the $BaPbO_3$ thin films with $a = b = 4.288 \pm 0.002$ Å and $c = 4.249 \pm 0.004$ Å correspond to an orthorhombic structure, consistent with the bulk values $a = b = 4.27$ Å and $c = 4.25$ Å [32]. The fact that the lattice constants of the $BaSnO_3$ and $BaPbO_3$ films coincide with the bulk values indicates that the electronic band structures depicted in Figs. 1(a) - 1(d) can be applied for the present thin films.

To reveal the evolution of the electronic band structure of $BaSn_{1-x}Pb_xO_3$ as a function of $x$, we evaluated the optical conductivity spectra $\sigma_1(\omega)$ of $BaSn_{1-x}Pb_xO_3$ thin films grown on both-side polished $SrTiO_3$ substrates via optical reflectance and transmittance measurements. By considering multiple reflections [33] within the sample and $SrTiO_3$ substrate, we extracted $\sigma_1(\omega)$ of $BaSn_{1-x}Pb_xO_3$ thin films (see Section III in SM [30]). Figure 2(a) shows the real part of $\sigma_1(\omega)$ of $BaSn_{1-x}Pb_xO_3$ for representative values of $x$ as a function of photon energy $\hbar\omega$ in the infrared region (in main panel) and the ultraviolet/visible (UV/VIS) region (inset), where $\hbar$ is the reduced Planck constant, which is equal to the Planck constant h divided by $2\pi$. $\sigma_1(\omega)$ in the infrared and UV/VIS regions were obtained from the fitting procedure using the RefFIT program [34,35] for the reflectance (Fig. S7 in SM [30]) and transmittance (Fig. S6 in SM [30]) data, respectively. The absorption edge for $x = 0$ is consistent with the band gap of bulk $BaSnO_3$ [28]. For the low $x$ with $x = 0$-$0.55$ (inset in Fig. 2(a)), the absorption edge systematically shifts toward the lower-energy region (from 3.1 eV for $x = 0$ to 1.73 eV for $x = 0.55$). According to the principle of band-gap engineering in semiconductors [36,37], *B*-site substitution of Sn by Pb is expected to cause hybridization between Sn-5*s* and Pb-6*s* orbitals in the conduction band. It is plausible that the band-gap narrowing at the Γ point illustrated in Figs. 1(a) and 1(b) occurs with increasing $x$ for $x < 0.55$. In addition to the signature of bandgap narrowing, the spectral weight (SW) of $\sigma_1(\omega)$ below 1 eV develops with increasing $x$ as presented in the main panel of Fig. 2(a), which indicates an appearance of metallic Fermi surfaces in Pb-rich $BaSn_{1-x}Pb_xO_3$. Especially, $\sigma_1(\omega)$ for $x = 1$ shows the so-called Drude response as observed in the earlier report [38], which is a hallmark of metals. Here, we emphasize that the value of $\sigma_1$ in the dc limit for $x = 1$ is consistent with the dc electrical conductivity in Fig. 3. A decrease of $\sigma_1$ below ~0.05 eV for $x = 0.55$, 0.85, and 0.90 comes from the presence of weak localization (WL) of conduction electrons, which is indeed observed in magnetotransport measurements as a positive magnetoconductance (MC) as discussed later (see Fig. 4).

More importantly, despite the continuous increase of SW in the low energy region above $x = 0.55$, its value for $x = 0.9$ is unexpectedly suppressed. To clarify such an anomalous feature, we constructed a contour plot of $\sigma_1$ as a function of $x$ (horizontal axis) and photon energy (vertical axis) in Fig. 2(b). A sudden suppression of $\sigma_1(\omega)$ for $x = 0.9$ (white arrow) is clearly exhibited although $\sigma_1(\omega)$ in the low-energy region basically increases with increasing $x$ because of the increase in the DOS around $E_F$. These results clearly reveal that an anomalous variation of the electronic structure occurs around $x = 0.9$. The plausible origin of the anomaly may be related to a small Fermi surface owing to a band touching, leading to a band inversion in Fig. 1(c) in BaSn$_{1-x}$Pb$_x$O$_3$.

An anomalous electronic band structure should have a significant impact on low-energy excitations, which can be seen in electrical transport properties. Therefore, we performed resistivity and magnetoconductance measurements in the series of BaSn$_{1-x}$Pb$_x$O$_3$ films. The longitudinal and Hall resistances were measured using the five-terminal method. The temperature $T$ dependence of the resistivity $\rho_{xx}(T)$ of the BaSn$_{1-x}$Pb$_x$O$_3$ films for various $x$ is presented in Fig. 3. Note that the value of the resistivity of the BaSn$_{1-x}$Pb$_x$O$_3$ films with $x = 0$ and 0.2 exceeded the measurement limit at room temperature, reflecting the insulating nature of the wide band-gap material. For $x = 0.42$ (purple) and 0.55 (navy) in the left panel, $\rho_{xx}(T)$ still exhibits insulating behavior (d$\rho_{xx}$/d$T < 0$) over the whole temperature range. For the larger $x$, an insulator-to-metal transition appears above $x = 0.74$. With further increase of $x$, the metallicity is improved up to $x = 0.85$. In contrast to the monotonic reduction of $\rho_{xx}$ in the left panel, we observed an inconsistent increase of $\rho_{xx}(T)$ at $x = 0.90$ (red) in the right panel. Finally, metallic behaviors appear for $x = 0.95$ and 1.0, in good agreement with an earlier report on bulk $BaPbO_3$ [38]. The comparable value of the resistivity compared with that of the bulk (~ $2 \times 10^{-4}$ $\Omega$cm at room temperature [38]) suggests that the metallic transport properties of the $BaPbO_3$ thin film originate from the semimetallic electronic band structure. Therefore, the specific band modification needs to be taken into account to explain both the overall systematic dependence of $\rho_{xx}$ with $x$ and the anomalous increase of that for $x = 0.9$, which we will discuss in a later part.

MC measurements are quite effective to detect band features through the spin dynamics of conducting electrons. Figure 4(a) displays the perpendicular magnetic field $B$ dependence of the sheet conductance $\sigma_{xx}(B)$ for the $BaSn_{1-x}Pb_xO_3$ films with $x = 0.55$ (navy), 0.74 (blue), 0.79 (green), 0.90 (red), and 1.0 (orange). Here, $\sigma_{xx}(B)$ is calculated by $R_{xx}^2(B)/(R_{xx}^2(B) + R_{yx}^2(B))$ with $R_{xx}$ and $R_{yx}$ being the sheet and Hall resistances, respectively. For clarity, the MC defined as $\Delta\sigma_{xx}(B) = \sigma_{xx}(B) - \sigma_{xx}(0)$ is shifted along the vertical axis. For the film with $x = 0.55$, we observed a positive MC with an order of magnitude of ~ $e^2/h$, characteristic of WL. Here, $e$ is the elementary charge. The occurrence of WL is consistent with the suppression of $\sigma_1(\omega)$ in the low-energy region shown in Fig. 2(a). In sharp contrast, for the films with $x > 0.74$, a sharp negative MC appears at lower fields below the characteristic magnetic field denoted by triangles in Fig. 4(a), followed by the positive MC due to WL at high fields. In the presence of a strong SOI, electron spins precess along the effective magnetic fields, leading to the destructive interference observed as weak antilocalization (WAL). Judging from the fact that the perovskite lattice of the BaSn$_{1-x}$Pb$_x$O$_3$ films is fully relaxed (see Fig. S2 [30]), the Rashba-type SOI owing to the band structural inversion asymmetry normal to the film plane is considered to play a dominant role [39]. The characteristic field reaches the maximum for $x = 0.90$. The transition from WL to WAL and the strong enhancement of the characteristic magnetic field originate from the increased contribution of the SOI owing to the band-gap narrowing [27]. To quantitatively evaluate the strength of the SOI in the BaSn$_{1-x}$Pb$_x$O$_3$ films, the WAL data were analyzed using the Hikami-Larkin-Nagaoka (HLN) model in two dimensions [30,40] with characteristic magnetic fields $B_{SO}$ and $B_\varphi$ as free parameters. Here, $B_{SO} = \hbar/4eL_{SO}^2$ and $B_\varphi = \hbar/4eL_\varphi^2$, where $L_{SO}$ and $L_\varphi$ are the spin-orbit scattering length and the phase coherent length, respectively. The black solid lines in Fig. 4(a) are the best fit using the HLN model, where we obtained a good agreement with the experimental data. WAL is observed only when $L_{SO}$ is shorter than $L_\varphi$ (*i.e.*, $B_{SO} > B_\varphi$) as shown in the inset of Fig. 4(b).

Figures 4(b) and 4(c) summarize the $x$ dependences of $B_{SO}$ and $B_\varphi$, and those of the electrical resistivity $\rho_{xx}$ at $T = 2$ K (green triangles) and SW of $\sigma_1(\omega)$ (orange circles) extracted from Figs. 3 and 2(a), respectively. The SW was evaluated from the energy integral of $\sigma_1$ from 0 to 1 eV. Based on these comprehensive data, we discuss the experimental verification of the electronic band structures depicted in Figs. 1(a) – 1(d). The band insulating states in Fig. 1(a) correspond to $0 < x < \sim 0.2$; the insulating behavior is demonstrated by the absence of a Drude response in $\sigma_1(\omega)$ and the high electrical resistivity. The metallic degenerate semiconducting states in Fig. 1(b) correspond to $\sim 0.2 < x < \sim 0.85$; the insulator-to-metal transition is consistently observed in the systematic decrease of $\rho_{xx}$ and increase of SW as shown in Fig. 4(c). The Hall coefficient $R_H$ in Fig.

S15(c) in SM [30], which is defined as $R_H = (R_{yx}/B)d$ with $d$ being the film thickness, shows a similar trend with the resistance (Fig. 4(c)), indicating a less conducting charge-carrier density at $x = 0.9$. We speculate that the cubic-to-orthorhombic structural phase transition at around $x = 0.5$ may be relevant to the drastic variation of the optical spectra such as a saturation of the absorption edge (white broken line in Fig. 2(b) and Fig. S12 in SM [30]) and the appearance of the low-energy excitations in $\sigma_1(\omega)$. It is likely that the conduction band of BaSnO$_3$ at the Γ point is proportionally shifted downward, with increasing the degree of hybridization of the Pb-6$s$ and Sn-5$s$ orbitals as shown in Figs. 1(b) and 1(c). The presence of a singular point in the optical and electrical conductivities (Fig. 4(c)) as well as the peak structure of the spin-orbit magnetic field (Fig. 4(b)) at around $x = 0.9$ strongly evidences an exceptionally low DOS with a narrow band gap [27], clearly demonstrating a band touching with a small Fermi surface as presented in Fig. 1(c). Note that the crystal structure of the characteristic BaSn$_{0.1}$Pb$_{0.9}$O$_3$ remains orthorhombic. The small finite conductivities remaining at this critical composition can be possibly ascribed to the contribution from the hole pockets at the $R$ and $M$ points above a certain Pb content (Fig. S14 in SM [30]) [22,25,26]. Finally, for $x = 1.0$, the Drude response and the metallic transport behavior were observed. The appearance of the specific features of the electronic band in the characteristic physical properties for $x = 0.9$ clearly exemplifies TPT in BaSn$_{1-x}$Pb$_x$O$_3$, directly revealing the nontrivial band structure of BaPbO$_3$ as depicted in Fig. 1(d).

In conclusion, we have investigated the optical conductivity and electrical magnetotransport properties of PLD-grown BaSn$_{1-x}$Pb$_x$O$_3$ thin films to exemplify TPT from the band insulator BaSnO$_3$ to the semimetal BaPbO$_3$. We have observed an insulator-to-metal transition in both the optical conductivity spectra and the temperature dependence of the resistivity. An anomalous suppression of the optical and electrical conductivities at a critical $x = 0.9$ constitutes persuasive evidence of a small Fermi surface, which directly links to the electronic band modification with a band inversion. This conclusion is also supported by the enhanced spin-orbit magnetic field extracted from the low-field magnetoconductance. The intriguing anomaly of the electronic structure associated with the band-gap modification presented in this study can be explored by more direct ways such as electrical transport measurements in mesoscopic wires [3] or in field-effect transistor configurations with electrostatic tuning of the Fermi level, and angle-resolved photoemission spectroscopy measurements [1,4-6]. Our findings provide a new platform for investigating topological features combined with a wide variety of physical properties inherent in perovskite oxide materials.

**Acknowledgments**

We thank Hidenori Takagi for fruitful discussions and Yuka Ikemoto and Taro Moriwaki for technical assistance. Far-infrared reflectivity measurements using a synchrotron radiation light source were performed at SPring-8 with the approval of the Japan Synchrotron Radiation Research Institute (2018B0073). This work was supported by Grants-in-Aid for Scientific Research (Grant No. 15H05699, No. JP15H05853, No. 16H05981) from the Japan Society for the Promotion of Science and CREST (Grant No. JPMJCR18T2), the Japan Science and Technology Agency. Chemical composition analysis was performed under the Inter-University Cooperative Research Program of the Institute for Materials Research, Tohoku University.

**Figure captions**

**FIG. 1.** (a)-(d) Schematics of electronic band structures at the $\Gamma$ point of BaSn$_{1-x}$Pb$_x$O$_3$ with Pb contents $x$ from (a) $x = 0$, (b) $x \sim 0.2$, (c) $x \sim 0.9$, and (d) $x = 1.0$. (e) $x$ dependence of the lattice parameters for the $c$ axis (red circles) and the in-plane axes (green squares). Dashed lines are guides to the eye. Inset: Crystal structure of the perovskite oxide BaSn$_{1-x}$Pb$_x$O$_3$.

**FIG. 2.** (a) Optical conductivity spectra $\sigma_1(\omega)$ of the BaSn$_{1-x}$Pb$_x$O$_3$ thin films with $x = 1$ (orange), 0.90 (red), 0.85 (green), 0.55 (navy), 0.20 (purple), and 0 (black) obtained from fitting the reflectance and transmittance data by using RefFIT program [30]. Note that the data for $x = 0$ in the infrared region are overlapped with those for $x = 0.20$. The circles at $\omega = 0$ represent values of dc conductivities obtained from transport measurements. The shaded area below 0.03 eV indicates the unmeasured region. Inset: $\sigma_1$ for $x = 0$, 0.2, and 0.55 in the UV/VIS region. (b) Contour mapping of $\sigma_1$ as a function of photon energy (vertical axis) and $x$ (horizontal axis), which was constructed by interpolating the values of $\sigma_1(\omega)$ between the adjacent $x$. The absorption edge in the UV/VIS region (white broken line), corresponding to the optical gap, continuously shifts to the lower energy region from $x = 0$ to 0.55, and saturates above $x = 0.55$.

**FIG. 3.** Temperature dependence of the resistivity $\rho_{xx}$ for the 18-20-nm BaSn$_{1-x}$Pb$_x$O$_3$ thin films with $x = 0.42$ (purple), 0.55 (navy), 0.74 (blue), 0.79 (green), and 0.85 (light green) in the left panel, and $x = 0.9$ (red), 0.92 (deep orange), 0.95 (yellow), and 1.0 (orange) in the right panel.

FIG. 4. (a) Magnetoconductance of the BaSn$_{1-x}$Pb$_x$O$_3$ thin films with $x$ = 0.55 (navy), 0.74 (blue), 0.79 (green), 0.90 (red), and 1.0 (orange). The black solid lines are the best fit using the HLN model. (b,c) $x$ dependences of (b) the characteristic magnetic fields $B_{SO}$ (red circles) and $B_\varphi$ (blue square) and (c) the resistivity $\rho_{xx}$ (left axis) and the SW (right axis).

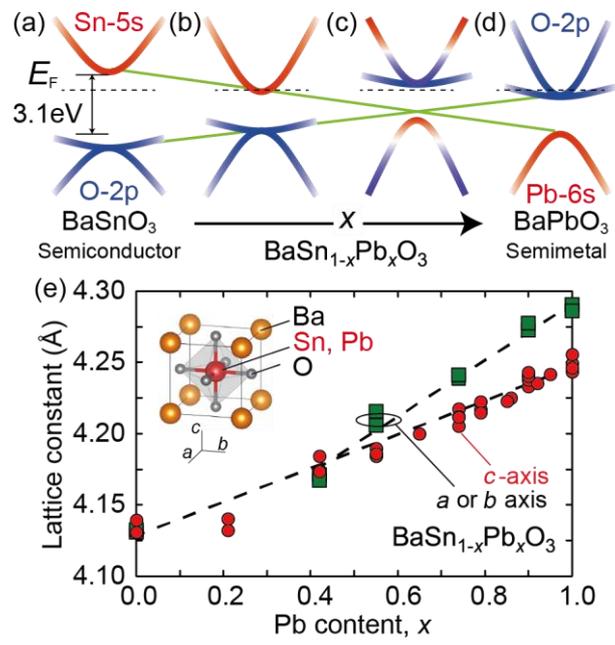

FIG. 1. J. Shiogai et al.

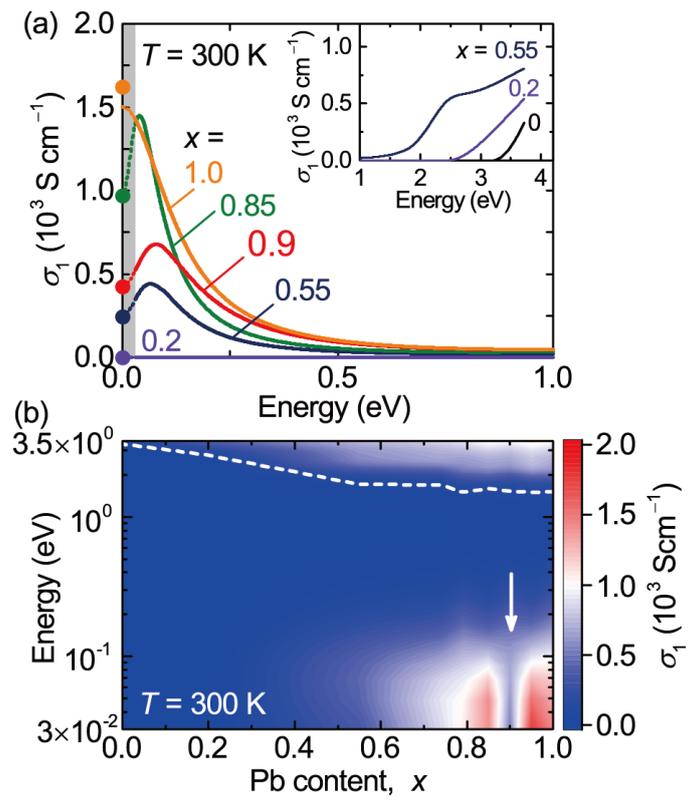

FIG. 2. J. Shiogai et al.

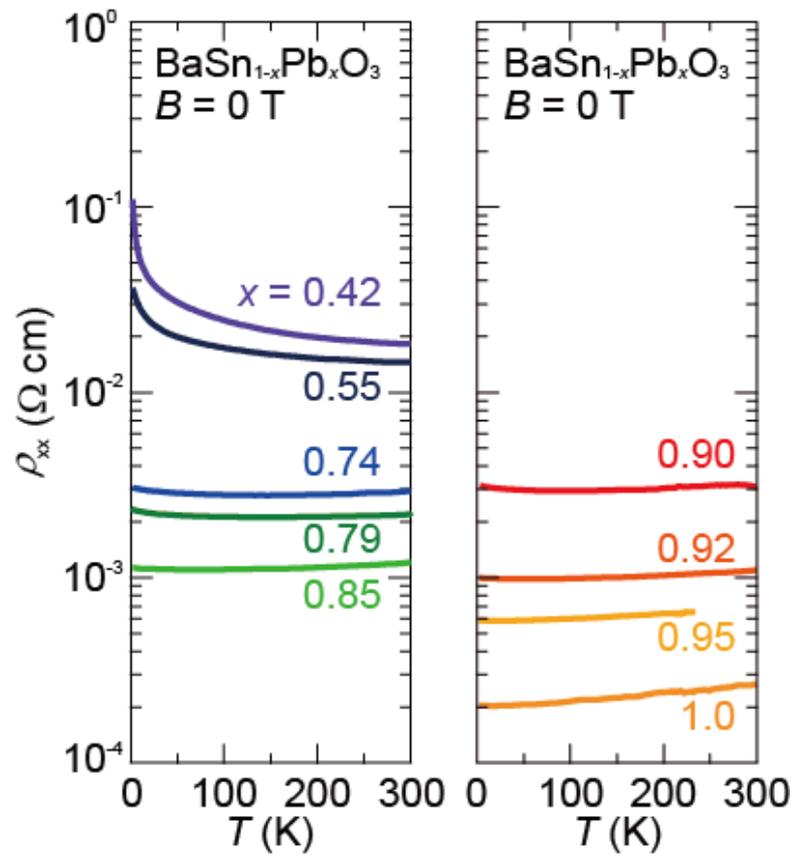

FIG. 3. J. Shiogai et al.

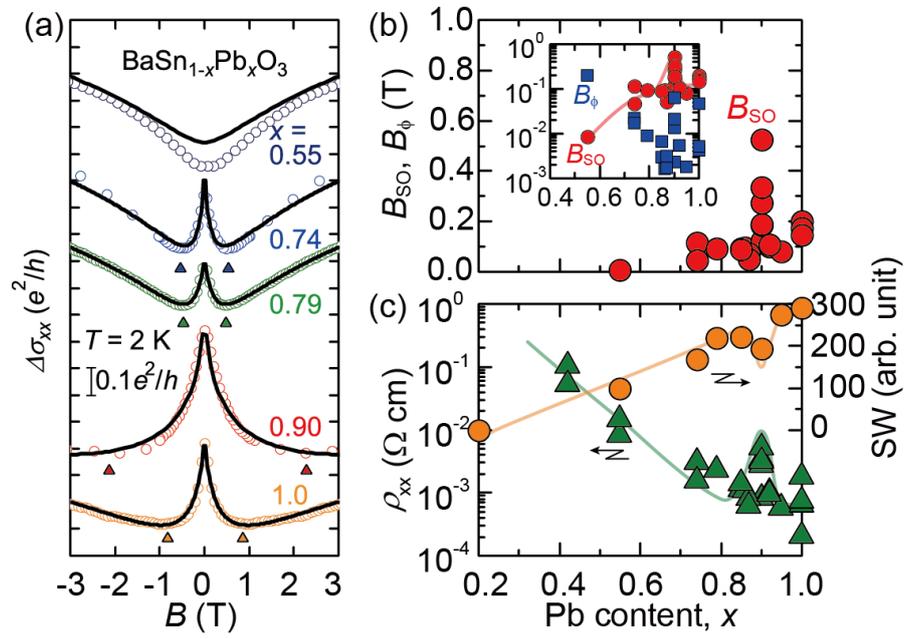

FIG. 4. J. Shiogai et al.